\documentclass[final]{IEEEtran}

\usepackage{amsmath,amsfonts,amssymb,amsopn}
\usepackage{graphicx}
\usepackage{xcolor}
\usepackage{cite}
\usepackage{xspace}
\usepackage{subfig} 
\usepackage{mathtools} 
\usepackage{url}
\usepackage{algorithm}
\usepackage{algpseudocode}

\usepackage{float}
\usepackage{xcolor}
\restylefloat{algorithm}
\floatname{algorithm}{\color{black}Procedure}

\def\NoNumber#1{{\def\alglinenumber##1{}\State #1}\addtocounter{ALG@line}{-1}} 

\ifCLASSOPTIONonecolumn
    
    \renewcommand{\baselinestretch}{2}
    \usepackage[margin=25mm]{geometry}
\else
    
\fi

\newcommand{\bv}[1]{\boldsymbol{#1}}
\newcommand{\dfn}{\triangleq}
\newcommand{\untsph}{\mathbb{S}^{2}} 

\newcommand{\shc}[3]{({#1})_{#2}^{#3}}
\newcommand{\lsph}{L^2(\untsph)}
\newcommand{\unit}[1]{\bv{\hat{#1}}}
\newcommand{\lsphL}[1]{{}_s\mathcal{H}_{#1}}

\newcommand{\intphi}{\int_{0}^{2\pi}}

\newcommand{\figref}[1]{Fig.\,\ref{#1}}

\newcommand{\secref}[1]{Section\,\ref{#1}}

\newcommand{\tilf}{\tilde f}
\newcommand{\fs}{\ensuremath{{}_s f}}

\newcommand{\ssod}{\mathfrak{S}^{L}_{s}}

\newcommand{\ylms}[2]{\ensuremath{{}_s Y_{#1}^{#2}}}

\makeatletter
\renewcommand\fs@ruled{\def\@fs@cfont{\bfseries}\let\@fs@capt\floatc@ruled
  \def\@fs@pre{{\color{black}\hrule height.8pt depth0pt \kern2pt}}%
  \def\@fs@post{{\color{black}\kern2pt\hrule\relax}}%
  \def\@fs@mid{{\kern2pt\color{black}\hrule\kern2pt}}%
  \let\@fs@iftopcapt\iftrue}
\makeatother

\DeclarePairedDelimiterX\innerp[2]{\langle}{\rangle}{#1,#2}

\newtheorem{remark}{Remark}

\graphicspath{{figs/},{pdfs/},{pdf/},{Figures/}}

\begin{document}
\title{An Optimal-Dimensionality Sampling for Spin-$s$ Functions on the Sphere}
\author{%
Usama Elahi,~\IEEEmembership{Student Member,~IEEE}, Zubair~Khalid,~\IEEEmembership{Member,~IEEE}, Rodney A. Kennedy,~\IEEEmembership{Fellow,~IEEE}, and Jason~D.~McEwen,~\IEEEmembership{Member,~IEEE}
\vspace*{-5mm}
\thanks{Usama Elahi and Rodney A. Kennedy are with the Research School of Engineering, College of Engineering and Computer Science, The Australian National
University, Canberra ACT 2601, Australia. Zubair Khalid is with the Department of Electrical Engineering, School of Science and Engineering, Lahore University of Management Sciences, Lahore 54792, Pakistan. Jason D. McEwen is with Mullard Space Science Laboratory, University College London, Surrey RH5 6NT, UK.
Usama Elahi and Rodney A. Kennedy are supported by the Australian Research Council’s Discovery Projects funding scheme (Project no.~DP170101897). Zubair Khalid is supported by Pakistan
HEC 2016-17 NRPU (Project no. 5925). Jason D. McEwen was partially supported by the Leverhulme Trust and the Engineering and Physical Sciences Research Council (grant numbers EP/M011852/1 and EP/M011089/1).
E-mail: usama.elahi@anu.edu.au, zubair.khalid@lums.edu.pk, rodney.kennedy@anu.edu.au, jason.mcewen@ucl.ac.uk.
}
}

\maketitle

\vspace*{-4mm}

\begin{abstract}
For the representation of spin-$s$ band-limited functions on the sphere, we propose a sampling scheme with optimal number of samples equal to the number of degrees of freedom of the function in harmonic space. In comparison to the existing sampling designs, which require ${\sim}2L^2$ samples for the representation of spin-$s$ functions band-limited at $L$, the proposed scheme requires $N_o=L^2-s^2$ samples for the accurate computation of the spin-$s$ spherical harmonic transform~($s$-SHT). For the proposed sampling scheme, we also develop a method to compute the $s$-SHT. We place the samples in our design scheme such that the matrices involved in the computation of $s$-SHT are well-conditioned. We also present a multi-pass $s$-SHT to improve the accuracy of the transform. We also show the proposed sampling design exhibits superior geometrical properties compared to existing equiangular and Gauss-Legendre sampling schemes, and enables accurate computation of the $s$-SHT corroborated through numerical experiments.
\end{abstract}
\vspace*{-1.5mm}
\begin{IEEEkeywords}
sampling, spherical harmonics, spin-$s$ functions, harmonic analysis, sphere
\end{IEEEkeywords}

\vspace*{-4mm}
\section{Introduction}

Spin functions (generally referred to as spin-$s$ functions) naturally arise in many applications including cosmology\cite{Zaldarriaga:1997}, astrophysics\cite{astro}, fluid dynamics\cite{Fluid-Fletcher}, global circulation modeling and models of stress propagation of earth\cite{Earth-ilbert}, to name a few. In particular, they play a pivotal role in the statistical studies of signals on the celestial sphere, such as cosmic microwave background (CMB) polarization and gravitational lensing\cite{Wiaux:2005:2,McEwen:2011,Huffenberger:2010}. In these applications, harmonic analysis is enabled through the spin\nobreakdash-$s$ spherical harmonic transform~($s$-SHT). Consequently, the ability to compute $s$-SHT of signal is of significant importance. It is desirable to design such sampling schemes that should require the minimum possible number of samples~(spatial dimensionality) for the accurate and efficient computation of the $s$-SHT of the signal and have superior geometrical properties~\cite{Driscoll:1994,McEwen:2011,mcewen:fsht,Kennedy-book:2013,Kostelec:2000:Spin,Khalid:2014,Huffenberger,Libsharp,Reinecke:2013,Usama:sampling}. For accurate computation of $s$-SHT of a band-limited spin-$s$ function with band-limit $L$~(formally defined in \secref{sec:models:b}) and spin $s$, the spatial dimensionality, denoted by $N_o$, is $(L^2-s^2)$ --- the number of degrees of freedom in harmonic space.

Discroll and Healy presented a sampling theorem on the sphere for an equiangular sampling grid of spatial dimensionality (asymptotically) ${\sim}4L^2$ for the exact reconstruction of band-limited scalar~($s=0$) functions on the sphere~\cite{Driscoll:1994}. For the same equiangular sampling grid, algorithms for the computation of $s$-SHT and signal reconstruction of band-limited spin-$s$ functions on the sphere $\mathbb{S}^2$ were developed in~\cite{Kostelec:2000:Spin}. A stable, fast and exact algorithm for equiangular sampling of spatial dimensionality ${\sim}4L^2$ for the evaluation of $s$-SHT of band-limited spin-$s$ functions with spin~($s=\pm2$) has also been proposed in \cite{Wiaux:2005:2}. These algorithms although enabling stable computation of  $s$-SHT, have large pre-computation and storage requirements. In order to reduce the pre-computation requirements, by exploiting the relationship between the Wigner $d$\nobreakdash-functions~\cite{Sakurai:1994,Kennedy-book:2013} and the spin-$s$ spherical harmonics, an accurate and exact algorithm to compute $s$-SHT using ${\sim}4L^2$ samples is proposed in~\cite{Huffenberger:2010}. To reduce the number of samples, \cite{mcewen:fsht} developed a sampling theorem requiring ${\sim}2L^2$ samples for the (theoretically)~exact and stable computation of $s$-SHT. The well-known Gauss-Legendre quadrature on the sphere\cite{Doroshkevich:2005} also supports exact computation of $s$-SHT using ${\sim}2L^2$ samples on the sphere. Recently, a library~(Libsharp)~\cite{Libsharp,Reinecke:2013} has been developed for the computation of $s$-SHT, where they reduce the number of samples of the Gauss-Legendre grid by approximately 30$\%$ by applying so-called polar optimization to reduce the number of samples around poles.

To the best of our knowledge, none of the existing sampling schemes requires fewer than ${\sim}2L^2$ samples for the accurate computation of $s$-SHT, well in excess of the optimal spatial dimensionality. In this context, we address the following questions in this work:
\begin{itemize}\vspace*{-0.5mm}
\item For band-limited spin-$s$ functions, how can we design a sampling scheme on the sphere with optimal spatial dimensionality of $N_o=L^2-s^2$?
\item Does the proposed scheme enable accurate computation of spin-$s$ spherical harmonic transform and exhibit superior geometric properties (such as mesh norm and mesh ratio) to existing schemes?
\end{itemize}

In addressing these questions, we organize the rest of paper as follows. We review the necessary mathematical background in Section \ref{sec:models}. The proposed sampling scheme and associated $s$-SHT are presented in section \ref{sec:samp}, where we also present a method to place samples on the sphere and develop a multi-pass $s$-SHT to improve the accuracy of the transform. We evaluate the numerical accuracy of the proposed transforms and analyse its geometrical properties in Section \ref{sec:anal}. %

\section{Preliminaries}
\label{sec:models}
\vspace*{-0.5mm}

\subsection{Spin-$s$ Functions on the Sphere}
\vspace*{-0.5mm}
We parameterize a point on the unit sphere $\mathbb{S}^2 \triangleq
\{\mathbf{u} \in \mathbb{R}^3 \colon |\mathbf{u}| = 1  \}$ as $\unit{u}\equiv\unit{u}(\theta,\phi)=(\sin\theta \cos\phi,\sin\theta \sin\phi,\cos\theta)$, where $\theta \in [0,\pi]$ and $\phi \in [0,2\pi)$ are the co-latitude and longitude respectively. The spin-$s$ functions on the sphere, denoted by $\fs\in\lsph$ with integer spin $s$, are defined by their behaviour under local rotation, that is, the spin-$s$ function transforms  as
$     \fs'(\theta,\phi)= e^{-is\gamma} \fs(\theta,\phi)$
under a local rotation $\gamma$. $\fs'$ is the rotated function obtained by rotating $\fs$ by $\gamma$ in the tangent plane at $(\theta,\phi)$.

\vspace*{-1mm}

\subsection{Spin-$s$ Spherical Harmonic Transform}
\label{sec:models:b}
\vspace*{-1mm}
The spin-$s$ spherical harmonic functions~(or spin weighted spherical harmonics), denoted by $\ylms{\ell}{m}$ for degree $\ell$, order $|m|\leq \ell$ and integer spin $|s|\leq \ell$, are defined as
\begin{align}\label{Eq:Y_spin}
\ylms{\ell}{m}(\theta,\phi) \dfn (-1)^s \sqrt{\frac{2\ell+1}{4\pi}} \, e^{im\phi} d_{\ell}^{m,-s}(\theta),
\end{align}
where $d_{\ell}^{m,m'}(\theta)$ denotes the Wigner-$d$ function~\cite{Sakurai:1994,Kennedy-book:2013}. Spin-$s$ spherical harmonics form a complete set of basis for spin-$s$ functions on the sphere and therefore any spin-$s$ function $\fs$ can be expanded as
\begin{align}\label{Eq:f_spin_expansion}
\fs(\theta,\phi) = \sum_{\ell=s}^{\infty}\sum_{m=-\ell}^{\ell} {\shc{\fs}{\ell}{m}}\, { \ylms{\ell}{m}(\theta,\phi)},
\end{align}
where ${\shc{\fs}{\ell}{m}}$ is the spin-$s$ spherical harmonic coefficient of degree $\ell$ and order $m$ and is given by the standard inner product for functions on the sphere~\cite{Kennedy-book:2013}:
\begin{align}\label{Eq:ssht}
{\shc{\fs}{\ell}{m}} =
\int_{\mathbb{S}^2}
\fs(\theta,\phi) \overline {\ylms{\ell}{m}(\theta,\phi)}
\,\sin\theta\, d\theta\, d\phi,
\end{align}
where $\overline{(\cdot)}$ denotes the complex conjugate operation. These coefficients form the spectral~(or harmonic) domain of the spin-$s$ function and the transformation of the spin-$s$ function to its harmonic coefficient given in \eqref{Eq:ssht} is referred to as spin-$s$ spherical harmonic transform~($s$-SHT).

The spin-$s$ function $\fs$ is said to be band-limited at $L$ if ${\shc{\fs}{\ell}{m}}=0$ for all $\ell\geq L$. Such band-limited spin-$s$ functions form a subspace $\lsphL{L}\subset \lsph $ of dimension $N_o$. For the representation of band-limited spin-$s$ function given in \eqref{Eq:f_spin_expansion}, the summation over degree $\ell$ is truncated to $L-1$.

\section{Optimal-Dimensionality Sampling and Spin-$s$ Spherical Harmonic Transform}
\label{sec:samp}
\vspace*{-0.5mm}

\subsection{Sampling Scheme}\label{sec:sampling_one}
\vspace*{-0.5mm}

We define an iso-latitude sampling scheme, denoted by $\ssod$, on the sphere as
\begin{multline}
\ssod = \big\{ (\theta_t,\frac{2\pi p }{2t+1} )~\big|~ t=|s|,|s|+1,\dotsc,L-1,\\[-2mm]  p = 0,1,\dotsc,2t \big\},
\end{multline}
which is comprised of $L-|s|$ iso-latitude rings of samples placed at $\theta_t\,, t=|s|,|s|+1,\dotsc,L-1$ with $2t+1$ equally spaced sampling points along $\phi$ in the ring placed at $\theta_t$. We discuss the location of iso-latitude rings later in this section. Since the sampling scheme takes $N_o$ sample points on the sphere in total, equal to the number of degrees of freedom of the spin-$s$ function band-limited at $L$, it is referred to as optimal-dimensionality sampling.

\subsection{Spin-$s$ Spherical Harmonic Transform -- Formulation}\label{sec:ssht:formulation}

We define the \emph{iso-latitude Fourier transform} of the signal $\fs(\theta,\phi)$ along $\phi$ given by

\begin{align}\label{Eq:gm_integral}
G_m(\theta) &\dfn \intphi \fs(\theta,\phi) e^{-im\phi} d\phi \nonumber \\
            &= (-1)^s \,  2\pi \sum_{\ell=\Delta}^{L-1}  \sqrt{\frac{2\ell+1}{4\pi}} \, \shc{\fs}{\ell}{m} \, d_{\ell}^{m,-s}(\theta),
\end{align}
where $\Delta=\max(|m|,|s|)$ and we have employed \eqref{Eq:Y_spin}, \eqref{Eq:f_spin_expansion} and orthogonality of complex exponentials in obtaining the second equality. Also define a vector of length $L-\Delta$ containing $G_m(\theta)$ evaluated at sample points of the proposed sampling scheme as
\begin{align}
{_s}\mathbf{g}_m &\dfn [G_m(\theta_{\Delta}),\,G_m(\theta_{\Delta+1}),\dotsc, G_m(\theta_{L-1})],
\end{align}
which can be equivalently expressed using \eqref{Eq:gm_integral} as
\begin{align}
_s\mathbf{g}_m  &= (-1)^s \, {_s}\mathbf{D}_m \,  _s\mathbf{f}_m
\label{Eq:gtof_inverse}.
\end{align}
where $_s\mathbf{D}_m \dfn$
\begin{equation}\label{Eq:Y_matrix}
\begin{small}\setlength{\arraycolsep}{1mm}
\begin{pmatrix}
   a_0 d_{\Delta}^{m,-s}(\theta_{\Delta}) & a_1 d_{\Delta+1}^{m,-s}(\theta_{\Delta}) & \cdots & a_{L-1}d_{L-1}^{m,-s}(\theta_{\Delta}) \\[4mm]
   a_0 d_{\Delta}^{m,-s}(\theta_{{\Delta}+1}) & a_1 d_{\Delta+1}^{m,-s}(\theta_{{\Delta}+1}) & \cdots & a_{L-1}d_{L-1}^{m,-s}(\theta_{{\Delta}+1}) \\
   \vdots  & \vdots  & \ddots & \vdots  \\
   a_0 d_{\Delta}^{m,-s}(\theta_{L-1}) & a_1 d_{\Delta+1}^{m,-s}(\theta_{L-1}) & \cdots & a_{L-1} d_{L-1}^{m,-s}(\theta_{L-1})
  \end{pmatrix}\end{small},
\end{equation}
with $a_{\beta}=\sqrt{\pi (2\beta+1)}$ and
\begin{equation}
_s\mathbf{f}_m  = \big[ \shc{\fs}{{\Delta}}{m},\,  \shc{\fs}{{\Delta}+1}{m},\,\dotsc,\, \shc{\fs}{L-1}{m}\big]^T,
\end{equation}
is a vector of spin-$s$ spherical harmonic coefficients of order $|m| < L$ and integer spin $s$.
\begin{remark}[Requirements for the Computation of $s$-SHT]\label{remark:transform}
The spherical harmonic coefficients contained in a vector $_s\mathbf{f}_m$ for each $|m|<L$ can be recovered by inverting the system in \eqref{Eq:gtof_inverse} provided (Requirement 1:) $_s\mathbf{g}_m$ is known and (Requirement 2:) ${_s}\mathbf{D}_m$ is invertible.
\end{remark}

\vspace*{-2mm}
\subsection{Spin-$s$ Spherical Harmonic Transform -- Computation}\label{sec:ssht:computation}

By changing the order of summations in \eqref{Eq:f_spin_expansion} and using \eqref{Eq:Y_spin} and \eqref{Eq:gm_integral}, we re-write \eqref{Eq:f_spin_expansion} for a band-limited spin-$s$ function as
\begin{align}\label{Eq:f_spin_expansion_2}
\fs(\theta,\phi) 
&= \frac{1}{2\pi} \sum_{m=-(L-1)}^{L-1} G_m(\theta) e^{im\phi},%
\end{align}
which indicates that $\fs(\theta,\phi)$ is a linear sum of $2L-1$ complex exponentials along $\phi$. Due to this fact, the existing iso-latitude schemes require $2L-1$ samples in {each} of the iso-latitude rings in order to compute $G_m(\theta_t)$ and consequently $_s\mathbf{g}_m$ correctly and serve the requirement 1~(see Remark~\ref{remark:transform}). However, if we know the spherical harmonic coefficients of all degrees~(and orders) greater than given $\Delta'$, we can remove their contributions from the signal which consequently reduces the band-limit along $\phi$ and enable the computation of $G_m(\theta_t)$ by taking an FFT over $2\Delta'+1$ samples (instead of $2L-1$ samples) in a ring placed at $\theta_t$. We further elaborate on this idea below.

We first compute the spin-$s$ spherical harmonic coefficients of orders $|m| = L-1$, which can be determined by computing ${}_s\mathbf{g}_{L-1}  = {}_s{G}_{L-1}(\theta_{L-1})$ and ${}_s\mathbf{g}_{-(L-1)}  = {}_s{G}_{-(L-1)}(\theta_{L-1})$ in \eqref{Eq:gtof_inverse} using an FFT over only one ring of $2L-1$ samples along $\phi$ placed at $\theta_{L-1}$. Once $\shc{\fs}{L-1}{L-1}$ is computed, we update the signal at the samples in the rings placed at $\theta_{t}$, $t=|s|,|s|+1,\dotsc,L-2$ as
\begin{equation}\label{Eq:signal_update}
{}_sf(\theta_t,\phi) \leftarrow {}_sf(\theta_t,\phi)-{}_s\tilf_{L-1}(\theta_t,\phi)
\end{equation}
where
\begin{align}\label{Eq:signal_subset}
_s\tilf_m(\theta,\phi) &\dfn \sum_{{\ell}=\Delta}^{L-1} \left\shc{\fs}{\ell}{m}  \ylms{\ell}{m}(\theta,\phi) + \shc{\fs}{\ell}{-m}  \ylms{\ell}{-m}(\theta,\phi) \right)\nonumber \\
&= \frac{1}{2\pi}\left(e^{im\phi} G_m(\theta) + e^{-im\phi} G_{-m}(\theta)\right)
\end{align}
denotes the part of the signal $\fs(\theta,\phi)$ composed of contribution of spherical harmonics of order $m$ and $-m$ and all degrees $\Delta \leq \ell\leq (L-1)$ for integer spin $s$. Once the signal is updated as given in \eqref{Eq:signal_update}, $2L-3$ samples are required instead of $2L-1$ to compute $G_m(\theta_t)$. For computing spin spherical harmonic coefficients of order $L-2$ and $-(L-2)$, we only need $2L-3$ samples along the $\phi$-ring placed at $\theta_{L-2}$. After computation, these can be used to update the signal at other sample positions. In this manner, we continue to evaluate the spin spherical harmonic coefficients for all degrees $\ell \geq |s|$ and all orders $|m|\le \ell$. This proposed $s$-SHT is summarized in the form of procedure below.


\begin{algorithm}
\caption{$s$-SHT}\label{Procedure:FSSHT}
\begin{algorithmic}[1]
\Require $\shc{\fs}{\ell}{m}$,\quad $\forall$  $|m| < L$, $\Delta\leq \ell <L$, given $\fs(\theta,\phi)$
\Procedure{Spin SHT}{$\fs(\theta,\phi)$}
\For{$m =L-1,L-2,\hdots,0$}
\State compute $_s\mathbf{g}_m$ and $_s\mathbf{g}_{-m}$ by evaluating $G_m(\theta_t)$
\NoNumber for $t = {\Delta}, {\Delta+1},\hdots,{L-1}$
\vspace{1.5mm}
\State compute $\mathbf{\fs}_m$ and $\mathbf{\fs}_{-m}$ by inverting \eqref{Eq:gtof_inverse}
\vspace{1.5mm}
%
%
\State update $\fs(\theta_t,\phi) \leftarrow \fs(\theta_t,\phi)-_s\tilf_m(\theta_t,\phi)$ for all
\NoNumber {$t=|s|,|s|+1,\hdots,\Delta-1$ and all associated \NoNumber sampling points along $\phi$}
\EndFor
\State \textbf{return} $\shc{\fs}{\ell}{m}$
\EndProcedure%
\end{algorithmic}
\end{algorithm}
\vspace{-3mm}

\subsection{Placement of Samples along Co-latitude}\label{sec:placement_samples}
We yet need to determine the sampling points along co-latitude for the placement of iso-latitude rings. To serve the requirement 2~(see Remark~\ref{remark:transform}), we propose a method, referred to as condition number minimization\cite{Khalid:2014}, to determine $L-|s|$ co-latitude sample points $\theta_t,\, t=\{|s|,|s|+1,\dotsc,L-1\}$ such that each matrix $_s\mathbf{D}_m$, given in \eqref{Eq:Y_matrix}, which depends on last $L-\Delta$ samples along co-latitude, is well-conditioned. Our method consider a set of $M\gg L$ equiangular samples~(with uniform measure along $\theta$) given by
%
$\Theta(M) = \big\{\frac{t\,\pi}{M+1} \big\}$, $t=1,2,\dotsc,M$
to choose sampling points $\theta_t,\, t=|s|,|s|+1,\dotsc,L-1$ using the following steps:

\begin{itemize}
\item Choose $\theta_{L}=\frac{{\pi}\lceil\frac{M}{2}\rceil}{M+1}$, that is, farthest from the poles in the set $\Theta(M)$, which is a natural choice for the ring of $2L-1$ samples along $\phi$.
\item For each $m=L-2,\,L-3,\,\dotsc,\,|s|$, choose $\theta_{m}$ from the set $\Theta(M)$ which minimizes the condition number of the matrix $_s\mathbf{D}_m$.
\end{itemize}
Such placement of iso-latitude rings ensures that each matrix $_s\mathbf{D}_m$ for at least each $|m|=|s|,|s|+1,\dotsc,L-1$ is well-conditioned and therefore enables the accurate computation of $s$-SHT. We note that the proposed condition number minimization method, although computationally intensive, is only required to be used once for each $s$ and band-limit $L$ to determine sample positions along co-latitude.
%
\begin{figure}[t]
    \centering
    \vspace{-6mm}
    \hspace{-2mm}
    \includegraphics[scale=0.4]{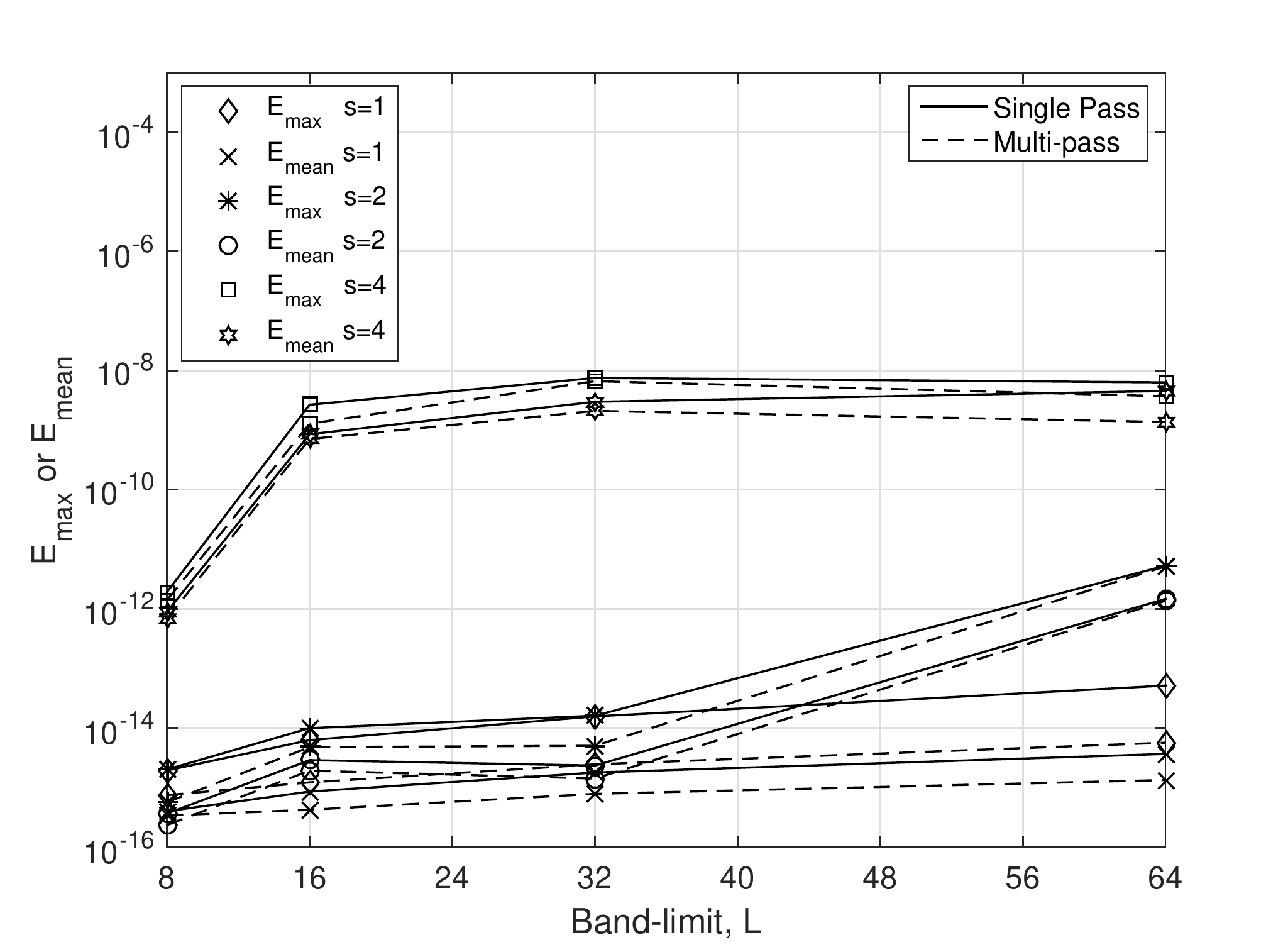}
    \vspace{-2mm}
    \caption{Numerical accuracy of the proposed $s$-SHT: the maximum error $E_\textrm{max}$ and the mean error $ E_\textrm{mean}$ for band-limits $L=(8,16,32,64)$ and integer spin $s=(1,2,4)$.}
    \label{fig:Ls2}
\end{figure}

\begin{figure*}[!ht]
    \vspace{-10mm}
    \centering
    \hspace{-4mm}
    \subfloat[]{
        \includegraphics[width=0.33\textwidth]{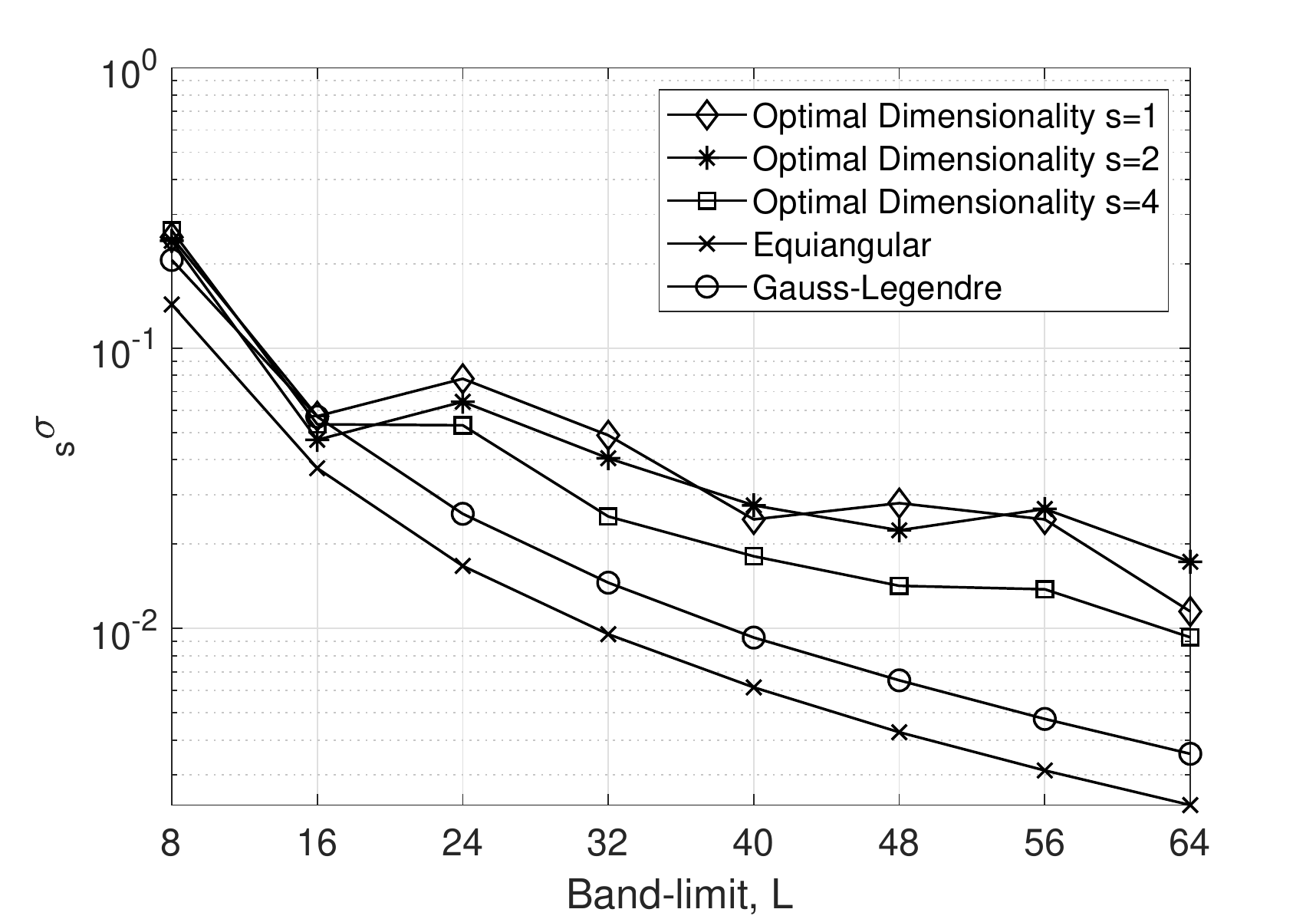}}\hfil
   \hspace{-3mm}
    \subfloat[]{
        \includegraphics[width=0.33\textwidth]{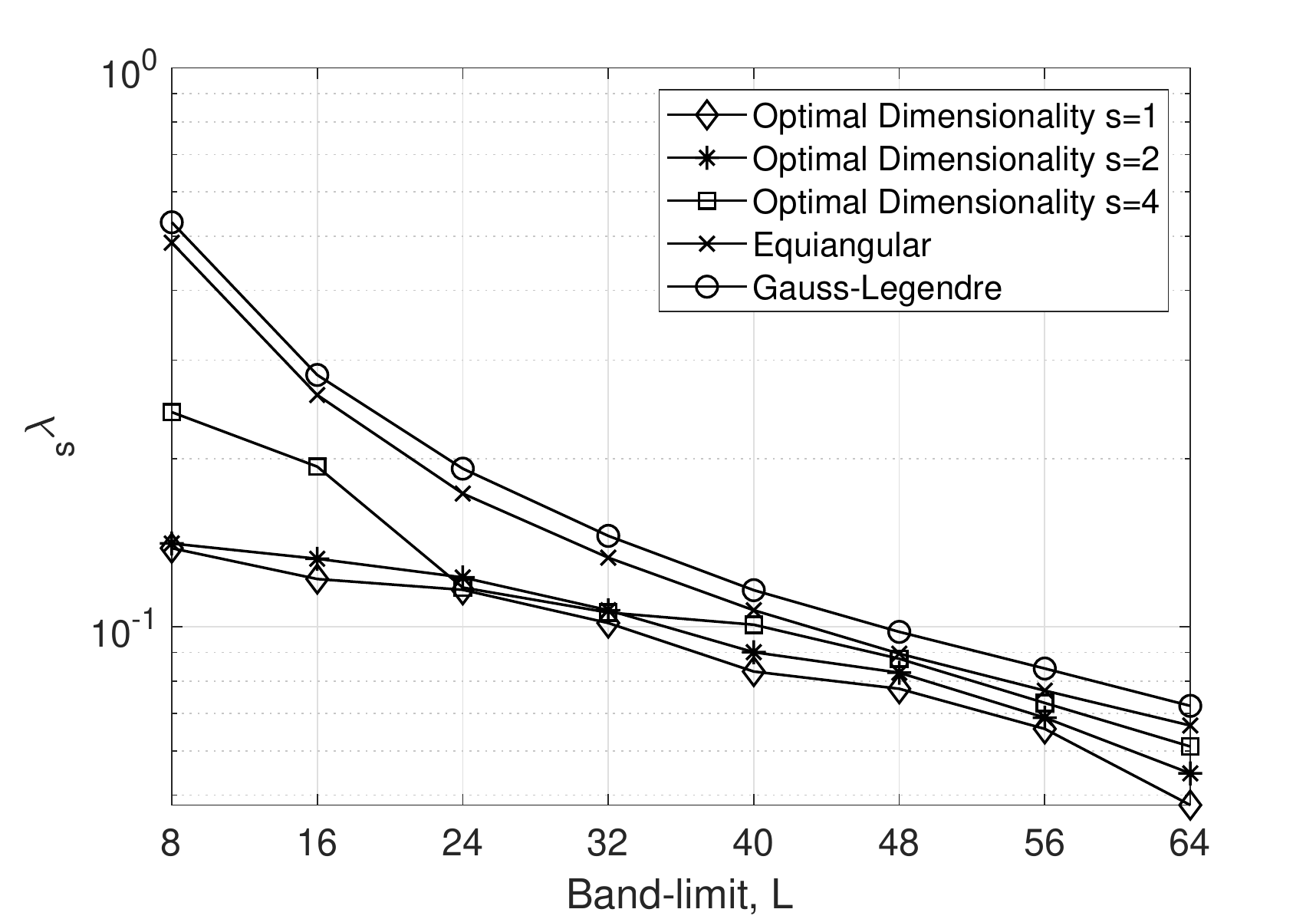}}\hfil
    \hspace{-4mm}
    \subfloat[]{
        \includegraphics[width=0.33\textwidth]{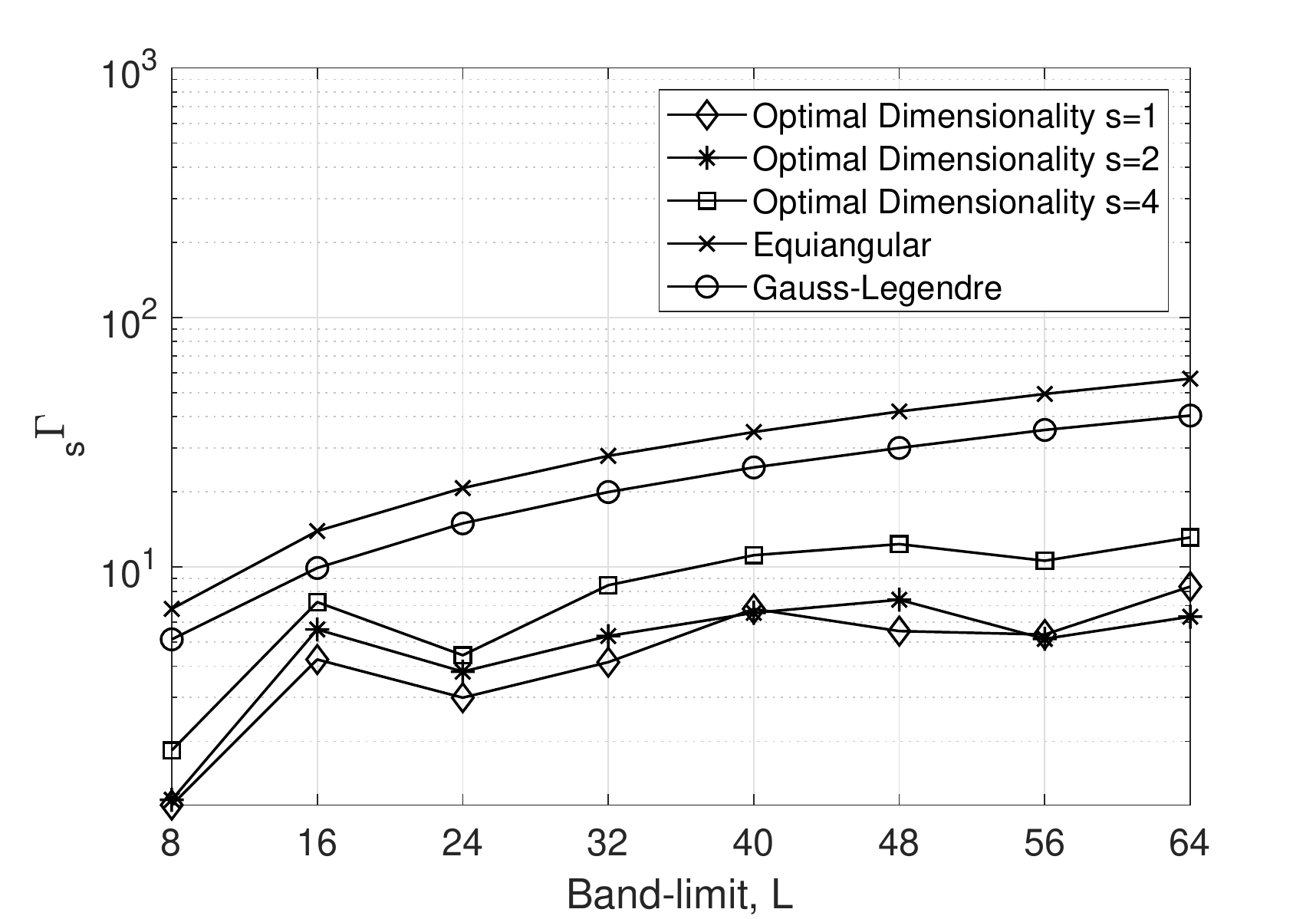}}\hfil
        \vspace{-2mm}
    \caption{The geometrical properties: (a) Minimum Geodesic Distance $_s\sigma(\zeta)$, (b) Mesh norm $_s\lambda(\zeta)$ and (c) Mesh Ratio $_s\Gamma(\zeta)$ for proposed, equiangular and Gauss-Legendre sampling schemes.\vspace*{-2mm}}
    \label{fig:geo_prop}
\end{figure*}

\vspace*{-4mm}
\subsection{Multi-pass $s$-SHT}

The $s$-SHT presented here computes the spherical harmonic coefficients for each order $m$ in a sequence $|m| = L-1, L-2, \dotsc, 0$. Since the error in the computation of $m$-th order coefficients contributes to the error in the computation of coefficients of order $m-1$, there tends to be error accumulation in the inherent sequential computation of $s$-SHT. We propose a multi-pass $s$-SHT which iteratively minimizes the building-up of error and therefore improves the accuracy of the $s$-SHT.

For a spin-$s$ function band-limited at $L$ and discretized over the sampling scheme $\ssod$, we compute its spherical harmonic coefficients, denoted by ${\shc{{}_s \breve{f}_1}{\ell}{m}}$ using proposed $s$-SHT and define the residual as
\begin{equation}
\label{Eq:residual}
r_k(\theta,\phi) = \fs(\theta,\phi) - \sum_{\ell=s}^{L-1}\sum_{m=-\ell}^{\ell} \shc{{}_s \breve{f}_k}{\ell}{m}  {}_sY_\ell^m(\theta,\phi)
\end{equation}
that is, an error between the signal $\fs$ and the signal synthesized from the spherical harmonic coefficients. Here $k=1,2,\dotsc,$ indicates the iteration number. We carry out $s$-SHT of the residual to compute its spherical harmonic coefficients, denoted by $\shc{{r_k}}{\ell}{m}$ and update $\shc{{}_s \breve{f}_k}{\ell}{m}$ as
\begin{align}
\label{Eq:update}
\shc{{}_s \breve{f}_{k+1}}{\ell}{m} = \shc{{}_s \breve{f}_k}{\ell}{m} + \shc{{r_k}}{\ell}{m}.
\end{align}
In multi-pass $s$-SHT, we propose to use \eqref{Eq:residual} and \eqref{Eq:update} iteratively to compute $\shc{{}_s \breve{f}_k}{\ell}{m}$ for $k=1,2,\dotsc$, until the following stopping criterion is met
\begin{align}
\label{Eq:criterion}
\max_{(\theta,\phi \in \ssod)} \big|r_{k+1}(\theta,\phi)\big| > \max_{(\theta,\phi \in \ssod)} \big|r_{k}(\theta,\phi)\big|.
\end{align}
Later, we illustrate, through numerical experiments, that the multi-pass $s$-SHT improves the accuracy over the (single-pass) $s$-SHT.
\vspace*{-4mm}
\section{Analysis}
\label{sec:anal}

\subsection{Numerical Accuracy}

In our experiment to evaluate the accuracy of the proposed spin-$s$ SHT, we generate the spin spherical harmonic coefficients $\shc{\fs_{\rm a}}{\ell}{m}$ of our test signal for $|s|<\ell<L$, $|m|\leq\ell$ with real and imaginary parts uniformly distributed in the interval $[-1,\,1]$ and then use \eqref{Eq:f_spin_expansion} to obtain the signal over samples of the proposed scheme $\ssod$. We then apply the proposed $s$-SHT and multi-pass $s$-SHT to reconstruct the spin-$s$ spherical harmonic coefficients denoted by  $\shc{\fs_{\rm r}}{\ell}{m}$. For each $L=8,16,32,64$ and $s = 1,2,4$, we repeat this experiment 10 times to obtain the average value of the maximum and mean errors, defined as
\begin{align}\label{Eq:exp2:errors:max}
E_\textrm{max} &\dfn \max\big|\fs_{\rm a}(\theta,\phi) - \fs_{\rm r}(\theta,\phi)\big|, \\ E_\textrm{mean} &\dfn \frac{1}{N_o} \sum_{(\theta,\phi)} \big|\fs_{\rm a}(\theta,\phi) - \fs_{\rm r}(\theta,\phi)\big| \label{Eq:exp2:errors:mean},
\end{align}
which are plotted in \figref{fig:Ls2} that illustrates that the proposed transforms enable accurate computation of $s$-SHT and multi-pass $s$-SHT improves the reconstruction accuracy. It is observed that the reconstruction errors grow with the increase in integer spin $s$, which is due to the fact that matrix ${}_sD_m$ defined in \eqref{Eq:Y_matrix} becomes ill-conditioned as $s$ increases irrespective of the placement of samples by the condition number minimization method.
\vspace*{-3.5mm}
\subsection{Geometrical Properties}

In this section, we compare the geometrical properties: minimum geodesic distance, mesh norm and mesh ratio for the proposed sampling scheme with those for the the existing equiangular~\cite{McEwen:2011} and Gauss-Legendre sampling schemes~\cite{Libsharp,Doroshkevich:2005} \footnote{Since these sampling schemes require different number of samples for the accurate reconstruction of spin-$s$ functions, we incorporate the sampling efficiency, denoted by $_sE_L$ and defined as the ratio of the dimensionality ($N_o$) of subspace formed by band-limited spin-$s$ functions to the number of samples required to accurately compute $s$-SHT, in our comparison.}.

For a set of sampling points on the sphere denoted by $\zeta$, the \emph{minimum geodesic distance}~(or the packing radius) is defined as the minimum distance between two points from a set of specific sampling points on the sphere, that is~\cite{sloan:2004}
\begin{align}\label{Eq:min_geo}
_s\sigma_n(\zeta) \dfn \frac{1}{_sE_L} \sigma(\zeta) =  \frac{1}{_sE_L}\,\min_{ \unit{u},\, \unit{v} \in  {\zeta} }   \,  {d_S}(\unit{u}, \unit{v}),
\end{align}
where ${d_S}(\unit{u}, \unit{v})=\cos^{-1}(\unit{u}.\unit{v})$ denotes the geodesic distance between $\unit{u}$ and $\unit{v}$~\cite{Kennedy-book:2013}. The \emph{mesh norm}~(or covering radius) is defined as the largest geodesic distance from any point $\unit{u} \in \untsph$ to its nearest point $\unit{v}\in\zeta$, given by~\cite{sloan:2009}
\begin{equation}\label{Eq:mesh_norm}
_s\lambda({\zeta}) \dfn \frac{1}{_sE_L} \,\max_{\unit{u} \in \mathbb{S}^2} \ \min_{\unit{v}\in \zeta} \ \  {d_S}(\unit{u},\unit{v}).
\end{equation}
The \emph{mesh ratio} for a sampling grid $\zeta$ is defined as~\cite{sloan:2009}
\begin{equation}\label{Eq:mesh_ratio}
_s\Gamma({\zeta})=\frac{2 _s\lambda(\zeta) }{_s\sigma_n(\zeta)} > 1,
\end{equation}
For a sampling scheme to facilitate the acquisition of samples, it is desirable to have larger minimum geodesic distance, smaller mesh norm and consequently smaller mesh ratio~\cite{Reimer:mesh,sloan:2009,sloan:2004,Usama:sampling}. We plot these properties for proposed sampling, equiangular sampling and Gauss-Legendre sampling schemes for band-limits $8 \le L \le 64$ and integer spin $s=1,2,4,$ in \figref{fig:geo_prop}, where it can observed that the proposed scheme exhibit superior geometrical properties primarily due to the dense sampling near the poles required by the existing schemes.
\vspace*{-3mm}
\section{Conclusions}\label{sec:conclusions}
For the accurate reconstruction of band-limited spin-$s$ functions on the sphere, we have proposed an optimal-dimensionality sampling scheme and developed a method to compute $s$-SHT associated with the proposed sampling scheme. We have placed the samples such that the linear system of equations involved in the computation of $s$-SHT are well-conditioned. For the accurate computation of $s$-SHT of the spin-$s$ function band-limited at $L$, the proposed sampling scheme requires optimal $N_o=L^2-s^2$ samples equal to the number of degrees of freedom of the signal in harmonic space. In comparison, the existing schemes require ${\sim}2L^2$ samples. We have also developed a multi-pass SHT that iteratively improves the accuracy of the transform, demonstrated the accuracy of the $s$-SHT and showed that the proposed sampling scheme is superior to existing schemes in terms of geometrical properties such as geodesic distance, mesh norm and mesh ratio.

\renewcommand{\baselinestretch}{1}


\end{document}